\documentclass[aps,prx,twocolumn,superscriptaddress,noshowpacs,preprintnumbers]{revtex4}

\usepackage[intlimits]{amsmath}
\usepackage{amsfonts}
\usepackage{graphics}
\usepackage{subfigure}
\usepackage[usenames]{color}
\usepackage{epstopdf,epsfig}
\usepackage{color}
\usepackage{indentfirst}
\usepackage[colorlinks=true,citecolor=blue]{hyperref}
\usepackage{ulem}
\usepackage{float}

\usepackage{graphicx}
\usepackage{amsmath}
\usepackage{amsfonts}
\usepackage[T1]{fontenc}
\usepackage{hyperref}
\hypersetup{
colorlinks=true,
citecolor=blue,
linkcolor=blue,
urlcolor=blue,
}
\usepackage{xcolor}

\makeatletter
\renewcommand{\p@subsection}{}
\renewcommand{\p@subsubsection}{}
\makeatother

\usepackage{braket}
\usepackage{bm}
\newcommand{\tr}[1]{\,\mathrm{tr}\left\lbrace  #1 \right\rbrace}
\newcommand{\beq}{\begin{equation}}
\newcommand{\eeq}{\end{equation}}

\begin{document}

\title{Mutual information for fermionic systems}

\author{Luca Lepori}
\affiliation{QSTAR and INO-CNR, Largo Enrico Fermi 2, 50125 Firenze, Italy.}
\affiliation{Dipartimento di Fisica, Universit\'a della Calabria, Arcavacata di Rende, I-87036 Cosenza, Italy}
\affiliation{I.N.F.N., Gruppo collegato di Cosenza, Arcavacata di Rende, I-87036 Cosenza, Italy}

\author{Simone Paganelli}
\affiliation{Dipartimento di Scienze Fisiche e Chimiche, Universit\`{a} dell'Aquila, via Vetoio,
I-67010 Coppito-L'Aquila, Italy}

\author{Fabio Franchini}
\affiliation{Ru{\dj}er Bo\v{s}kovi\'c Institute Bijen\v{c}ka cesta 54, 10000 Zagreb, Croatia}

\author{Andrea Trombettoni}
\affiliation{Department of Physics, University of Trieste, Strada Costiera 11, I-34151 Trieste, Italy}
\affiliation{SISSA and I.N.F.N., via Bonomea 265, I-34136 Trieste, Italy}
\affiliation{CNR-IOM DEMOCRITOS Simulation Center, 
Via Bonomea 265, I-34136 Trieste, Italy}

\preprint{RBI-ThPhys-2020-17}

\begin{abstract}
  We study the behavior of the mutual information (MI) 
  in various quadratic fermionic chains, with and without pairing terms
  and both with short- and long-range hoppings.
  The models considered include the short-range {\color{black} limit and
    long-range versions of the Kitaev model as well}, 
  and also cases in which the area law for the
  entanglement entropy is -- logarithmically or non-logarithmically -- violated.
  {\color{black}In all cases surveyed, when} the area law is violated at most logarithmically,
  the MI is a monotonically increasing function of the conformal
  four-point ratio $x$.
  Where non-logarithmic violations of the area law are present, 
  non-monotonic features can be observed {\color{black}in the MI and the
  	four-point ratio, as well as other natural combinations of the parameters, is found not to be
  	sufficient to capture the whole structure of the MI with a collapse onto a single curve. 
    We interpret this behavior as a sign that the structure of peaks is related to a non-universal spatial configuration of Bell pairs.}
  For the model exhibiting a perfect volume law, the MI vanishes identically. 
  For the Kitaev model {\color{black}
    the MI is vanishing for $x \to 0$} and {\color{black}it remains zero up to a finite $x$ in the gapped case.}
   {\color{black}In general, a larger range of the pairing corresponds to a reduction of the MI at small $x$.}
  A discussion of the comparison
  with the results obtained by the AdS/CFT correspondence
  in the strong coupling limit is presented.
\end{abstract}

\maketitle

\section{Introduction}
Over the years, information theory has provided valuable tools to analyze
and interpret many-body quantum systems. The entanglement entropy (EE),
as measured by the von Neumann and Rènyi entropies, 
has quickly
established itself as a standard benchmark that allows to discriminate
different phases of matter and properties escaping traditional paradigms.
EE provides relevant knowledge 
such as how much classical information is necessary for a faithful
representation of a quantum system or as an efficient witness of critical
behavior \cite{JPA2009,JSTAT2015}.

The EEs are bipartite measures, since they assess how much information is
shared between two complementary subsystems of a system in a pure state
(that is, with no intrinsic entropy as a whole). This information
cannot be accessed by measurements on one part alone, as it
reflects correlations between the two subsystems.
Since correlations lie mostly on the boundaries between the subsystems,
the area law, at most with logarithmic corrections, is typically obeyed
\cite{Eisert10}.

To access additional information on the structure of multipartite
entanglement
and quantum correlations, multipartite measures have also been introduced.
One such measure, although not an entanglement estimator, is the
so-called {\it mutual information} (MI). The MI is typically tripartite.
Starting with a system in a pure state, one indeed divides the system
into two (non-overlapping)
subsystems $A$ and $B$ and their complement $C \equiv\overline{A\cup B}$.
The MI measures how much information on $A$ we can obtain by measuring $B$
(or vice-versa). Thus, the MI can be considered as a kind
of two-point function, while, from this point of view, the EEs are essentially one-point functions.
Note, however, that an alternative point of view is to start
with a system made only of $A$ and $B$
in a mixed state obtained by tracing out $C$
and then to calculate the MI between $A$ and its
complement $B=\overline{A}$. Comparing the two points of view,
the latter can be considered as the result of removing $C$ and
letting $A$ and $B$ be "immersed" in the bath constituted by $C$.

Taking this into account,
in this work we will take the former approach and only consider
pure tripartite systems, initializing
the system under consideration in its (pure) ground state
\begin{equation}
\rho=\ket{\psi_{0}}\bra{\psi_{0}}.
\end{equation}
We then divide the lattice in the three disjoint parts $A$,
$B$ and their complement $C=\overline{A\cup B}$.
The reduced density matrix of each subsystem is given by
$\rho_{A}=tr_{B\cup C}\left\{ \rho\right\}$, 
$\rho_{B}=tr_{A\cup C}\left\{ \rho\right\}$,  
$\rho_{AB}=tr_{C}\left\{ \rho\right\}$, 
and $\rho_{C}=tr_{A\cup B}\left\{ \rho\right\}$, respectively. 
Starting from the von Neumann entropy of the subsystem $j$, defined as
\begin{equation}
S_{j}=-\tr{\rho_{j}\ln\rho_{j}},
\end{equation}
(where $\rho_j$ is the reduced density matrix of the subsystem $j$),
the MI between $A$ and $B$, denoted by $I_{(A,B)}$, is given by
\begin{equation}
I_{\left(A,B\right)}=S_{A}+S_{B}-S_{AB} \,.
\end{equation}

The MI is positive and symmetric in $A$ and $B$.
If $A$ and $B$ are complementary (that is $A \cup B$ equals the
entire system) and the whole system is prepared into a pure state, then
$S_{AB}\equiv S_{A \cup B}=0$ and $ I_{(A,B)}= 2 S_A = 2 S_B$,
thus the MI reduces to the von Neumann entropy.
The MI satisfies an area law \cite{wolf08} and
provides an upper bound to any two-point correlation function computed
between $A$ and $B$:
\begin{equation}
  I_{(A,B)} \ge {\Big( \langle {\cal O}_A \otimes {\cal O}_B \rangle 
  - \langle {\cal O}_A \rangle \langle {\cal O}_B \rangle \Big)^2 \over
  2 \left| {\cal O}_A \right|^2 \left| {\cal O}_B \right|^2 } \,.
\end{equation}
Thus, the MI vanishes when no correlation exists between $A$ and $B$. 

In the general case, $I_{(A,B)}$ is independent from UV cut-off parameters 
(since the divergences of the individual regions cancel over
those of their union) and this is one of the properties that renders it more
flexible over the entanglement entropy,
for which it is sometime complicated to disentangle the physical
from the cut-off contributions.
Another limitation of the EE is that it makes sense as a measure only
if the system is prepared in a pure state, while, as we remarked,
the MI applies also to mixed state systems.

The MI can be considered as a kind of two-point function,
since it depends on the relative position of regions $A$ and $B$. 
The MI is not a proper entanglement measure. 
However it is a measure of correlation between subsystems of a quantum system, and
it quantifies the amount of information 
shared between the two regions, providing a quantum counterpart
of the Shannon MI, which in general
quantifies in suitable units the amount of information
obtained about a random variable via the measurement of another
random variable \cite{Mac}.

The MI has been studied in several prototypical settings,
starting from the Ising model \cite{um12,lau13} and
critical chains \cite{alcaraz13,alcaraz14,stephan14},
in systems at finite temperature
\cite{melko10,singh11,wilms11,iaconis13,bernigau15} and
out of equilibrium \cite{growth,eisler14,asplung14,alba17,gruber20,maity20}.
Moreover, it has been used as a test for several phenomena,
for instance in disordered systems \cite{getelina16,ruggiero16,detommasi17} and in the presence of spontaneous symmetry breaking \cite{hamma16}, and
as well in the holographic settings
\cite{headrick2010,vilaplana11,allais12,fischler13}.
In one-dimensional (1D) systems described by a conformal field theory (CFT), 
several results have been obtained for the MI calculated with respect
to the Shannon entropies (that is, basis-dependent entropies)
\cite{alcaraz13,alcaraz14,stephan14}, while exact results
in lattice systems have been obtained
only in special cases \cite{brightmore19}. We refer to \cite{casini2009}
for a review on the EE in free quantum field theory with disjoint
intervals.

A general prediction for the behavior of the MI comes from the
anti-de Sitter (AdS)/CFT-correspondence, in the strong-coupling (large conformal charge $c$)
limit, where the MI has been found to behave as \cite{headrick2010,vilaplana11}
\begin{equation}
  I_{(A,B)}(x) = \begin{cases} 0\,,\quad & x < 1/2 \\ \frac{c}{3}
    \log \frac{x}{1-x} \,,\quad & x \geq 1/2\end{cases} \,,
\label{formula}
\end{equation}
as a function of the conformal four-point ratio $x$, defined as
\begin{equation} \label{eqn:cfpr}
x=\frac{l^{2}}{\left(l+d\right)^{2}} \, ,
\end{equation}
where $l$ is the length (i.e., the number of sites)
of the two subsystems $A$ and $B$ (taken to be the same size)
and, assuming $A$ and $B$ 
both simply connected. In
Eq. \eqref{eqn:cfpr}, $d$ is a distance between the two subsystems,
being the minimum number of sites between two points belonging to $A$ and $B$.
The four-point ratio is the traditional quantity used in CFT analysis \cite{DiFrancesco} used to encode the position of the four edges of the two subsystems.

Most of the studies of MI have focused so far mostly on systems with
short-range couplings/interactions. In this respect it would be interesting
to compare results for long-range systems with the corresponding
short-range findings, as motivated by recent results for quantum systems
with long-range couplings where the effects of long-rangedness on critical
properties, quantum dynamics and quantum entanglement properties have
been studied
\cite{celardo15,gong16,defenu16,defenu17,igloi18,blass18,defenu18,lerose19,pappalardi19,lerose20}.
One of the models in which the features induced by long-range terms have been
compared with their short-range counterparts 
is the long-range Kitaev model
\cite{vodola14,vodola2016,Buyskikh16,vanregemortel16,lepori16,dutta17,lepori17,alecce17,defenu19,defenu19bis,uhrich20}, where the pairing
term present in the short-range chain
$\sim \Delta c_i^\dag c_{i+1}^\dag$ \cite{kitaev} has the form 
$\sim \Delta_{ij} c_i^\dag c_{j}^\dag$.

Our goal in the present paper is fourfold: {\it i)} to compare the
short-range Kitaev model with the tight-binding chain to highlight the effect
of the pairing term on the MI; {\it ii)} to study how short-range results
for the MI get modified in the presence of long-range terms,
analyzing the MI in several prototypical 1D chains with long-range hoppings
and pairings; {\it iii)} to discuss the dependence of the MI
on the {\color{black} physical parameters, and in particular on the} conformal four-point ratio $x$, in the various models, including the
Kitaev model; {\it iv)} to probe the, possibly different, behavior of
the MI for systems displaying area law or its (non-logarithmic)
violation for the EE. We will consider quadratic and translational invariant
fermionic chains, allowing for the study of the MI for large system sizes
\cite{peschel2011}. {\color{black}Before moving on with the presentation
  we observe that in general the conformal four-point ratio is not necessarily the correct quantity to use in the study of the MI, especially in gapped systems.} {\color{black} It is anyway a useful combination of parameters that can be employed to compactly express the obtained results. In the following we will use $x$, {\color{black}as well as other parameters such} as $l/d$ according to what we find  more convenient in the different cases, commenting on other possible choices of combinations.
}

Regarding points {\it i)} and {\it iii)}, we observe
that although the models considered are
\textcolor{black}{not in the 
regime of validity for the AdS/CFT result}
(\ref{formula}), we aim at testing {\color{black} whether at least there is}
a qualitative agreement with it.
We will show that, when the area law is violated more
than logarithmically, the MI develops non-monotonic features,
with peaks for small $x$, in contrast with (\ref{formula}). We will also argue
that the four-point ratio is not
sufficient to capture the structure of
MI for systems with volume law of EE,
indicating an incompatibility of these cases with the assumption of conformal invariance and thus, the failure of both the CFT and holographic predictions.

The plan of the paper is the following.
In Section \ref{sec:models} we introduce the models we are going to consider, {\color{black}which comprise short and long-range, with and without pairing terms}.
Section \ref{sec:sr} is devoted to the short-range models,
with the tight-binding model considered in Section \ref{subsec:tb}
and the Kitaev chain in \ref{subsec:kit}. Long-range models are considered
in Sections \ref{sec:tb} and \ref{sec:pairing}: in the former the long-range
terms are the hopping ones, while in the latter we study the effect
of long-range pairing terms. In Section \ref{sec:tb}
we consider different forms of hoppings, {\color{black}which are known to generate different
phenomenologies in} the EE scaling: hoppings decaying as a power-law
exhibiting area law for the EE are considered in Section \ref{sec:tb}, while
the MI for model with fractal Fermi surface having a scaling of EE
intermediate between area and volume laws is discussed in Section \ref{sec:fractal}.
Models featuring EE volume law are studied in Section \ref{subsec:volume}, where
we consider {\color{black}both} a long-range power-law model with a space-dependent phase in the hopping in Section \ref{sec:LRPLPhase} and models with selective hoppings {\color{black}in Section \ref{sec:selhop}. Here} the hopping is chosen in a way to reproduce a state with the maximum number of Bell pairs, therefore giving EE volume law. We focus in particular on the model in which each site is coupled by an hopping term to the most distant site, a model introduced in \cite{gori15} and to which we refer to as the "antipodal" model. Deviations from the antipodal model are as well investigated. Section \ref{sec:pairing} considers the case of long-range pairings, both with short-range hoppings (Section \ref{sec:LRpairing}) and long-range hoppings (Section \ref{sec:LRHopPairing}). The use of MI to detect quantum phase
transitions in such models is studied.
Final comments and conclusions drawn from our results are collected in Section \ref{sec:conclusions}.

\section{The models}
In this paper, we consider 1D fermionic models of the form
\label{sec:models}
\begin{equation}\label{eqn:gen}
 H = H_H + H_P ,
\end{equation}
where $H_H$ is the hopping part and $H_P$ the pairing part of the Hamiltonian.  Sites are labeled by indices $i,j=1,\cdots,N_s$ and $N_S$ is the number of
sites. The fermionic operators creating and destroying a fermion {\color{black} on} site $i$ are denoted
by $c_i$ and $c_i^{\dagger}$, respectively. The filling $f$, {\color{black}which is also the number of particles per site, is defined as $f=N_T/N_S$, where $N_T=\sum_{i=1}^{N_S} c_i^{\dagger} c_i$ is the total
particle number operator.}
As usual, the filling can be fixed directly or via the introduction of a chemical potential
$\mu$, amounting to fixing the number of fermions from the Hamiltonian $H-\mu N_T$.
We will consider as well translational invariant models,
where, unless explicitly stated, periodic boundary conditions
(PBC) are imposed {\color{black}(antiperiodic boundary conditions will be instead imposed when long-range pairing terms will be considered)}:
\begin{equation}
  c_{i+N_S} \equiv c_i.
  \label{eqn:PBC}
\end{equation}  

The hopping part $H_H$ of the Hamiltonian (\ref{eqn:gen}) reads in general
\begin{equation}\label{eqn:Hamiltonian}
 H_H = -\sum_{i,j=1}^{N_S} t_{i,j} c_{i}^{\dagger}c_{j} + \mathrm{H.c.},
\end{equation}
where $t_{i,j}$ is the hopping amplitude among sites $i$ and $j$.
Several form of $t_{i,j}$ will be considered:
\begin{itemize}

\item[{\it a)}] nearest-neighbor, with $t_{i,j}\neq 0$ for  $j=i\pm 1$ and vanishing otherwise;

\item[{\it b)}] selective hopping, with $t_{i,j}$ constant and non-vanishing 
  if the distance between $i$ and $j$ is in an assigned interval of values. For instance, the antipodal model has $t_{i,j} \neq 0$ only if $|i-j|=N_S/2$ (with $N_S$ even).

\item[{\it c)}] long-range with
\begin{equation}
    \label{LR_alpha}
    t_{i,j} \propto \frac{1}{|i-j|_p^{\alpha}} \,,
\end{equation}
where the distance $|\cdot|_{p}$, due to PBC, is defined as 
\begin{equation}
|i-j|_{p}=\min\left(|i-j|, N_S-|i-j|\right) \,.
\label{definition_dist}
\end{equation}
Notice that while in Sections \ref{subsec:tb} and \ref{sec:tb}
the prefactor not written in (\ref{LR_alpha}) is considered real,
in the model discussed in Section \ref{sec:LRPLPhase} is complex.
  
\end{itemize}

The power-law exponent Eq. $\alpha$ in (\ref{LR_alpha}) is such that
for $\alpha \to \infty$ the short-range limit is retrieved. Moreover,
if $\alpha > d=1$ (more in general $\alpha$ larger than the dimension $d$ of the lattice)
then the energy is extensive \cite{libro}. When $\alpha>1$, in statistical mechanics
models one can find a value of $\alpha$, often denoted by $\alpha^*$, such that
for $\alpha > \alpha^*$ the critical behavior is the one of the short-range model \cite{sak1973},
although of course non-universal quantities may depend on $\alpha$ (see a review
in \cite{review}). One refers often to the range $\alpha \le 1$ as the "strong" long-range" region 
and to $1<\alpha < {\color{black} \alpha^* = 2}$ {\color{black}(for 1D chains)} as the "weak" long-range region, but in this paper, not being
crucially focused on critical properties, we will not make such distinction, making generic
reference to the law (\ref{LR_alpha}) as a long-range, power-law decay.
An important point emerging in the study of lattice models
with long-range interactions is that critical properties change with the exponent
$\alpha$ at a fixed dimension of the lattice in which interactions are long-range \cite{review}. So, from this
point of view, changing the dimension of the lattice (e.g., considering two-dimensional lattices), although
interesting, is expected not to qualitatively change the properties we are going to discuss. 

Since we are adopting PBC, the eigenfunctions of the matrix $-t_{i,j}$ are plane waves
$\psi_{k}(j)\propto e^{ikj}$, with
$k=\frac{2\pi}{N_S} n_k$ and $n_k=-N_S/2,\ldots,N_S/2-1$, again assuming
$N_S$ even (the lattice spacing is set for simplicity $\equiv 1$).
Therefore the hopping Hamiltonian (\ref{eqn:Hamiltonian})
can be readily diagonalized as $H_H=\sum_k \epsilon_k c_k^{\dagger} c_k$, with $c_k$ the Fourier transform
of $c_i$ and $\epsilon_k$ depending of course on the specific form of the $t_{i,j}$. In Section
\ref{sec:fractal} a model with fractal Fermi surface is studied, and there
the form of $\epsilon_k$ will be directly given without
explicitly assigning the hoppings $t_{i,j}$.

For the hopping Hamiltonian (\ref{eqn:Hamiltonian}) the EE of the
subsystem $A$ having $l$ sites is given by \cite{peschel2011}
\begin{equation}
S_{A}=-\sum_{\gamma=1}^{l} \left[ \left( 1-C_\gamma \right)
\ln{\left( 1-C_\gamma \right)} + C_\gamma \ln{C_\gamma} \right] \,.
\label{res}
\end{equation}
In (\ref{res}) the $C_\gamma$ 
are the $l$ eigenvalues of the correlation matrix 
\begin{equation}
C_{ij} =\langle \Psi \vert c_i^\dag c_j \vert \Psi \rangle \,,
\label{correl_intr}
\end{equation} 
with $i,j=1,\cdots,l$ being the sites belonging to $A$ and $\Psi$ the
ground-state of the fermionic system.

To calculate the MI, we calculate the EE of two systems
$A$ and $B$ having each $l$ sites,
then the EE of $A\cup B$, and finally the MI using (\ref{formula}).
In this way we can readily calculate and analyze the MI, and express it in term
of the four-point ratio where $d$ is the (minimal) distance between $A$ and
$B$. The results we are going to present
in the following are obtained using PBC, in order to treat a larger number of sites, however
we have also checked that open boundary conditions do not qualitatively alter the outcomes.

Let us now discuss the pairing term $H_P$ in the total Hamiltonian
(\ref{eqn:gen}):
\begin{equation}\label{eqn:Hamiltonian_P}
  H_P = \sum_{i,j=1}^{N_S} \Delta_{i,j} c_{i}^{\dagger} c_{j}^{\dagger} + \mathrm{h.c.}
  \,.
\end{equation}
When both the pairing term $\Delta_{i,j}$ and the hopping term
are nearest-neighbor ($\neq 0$ only if $j=i\pm 1$ and zero otherwise),
then one has the Kitaev
model \cite{kitaev}. Long-range in the pairings will be as well introduced:
\begin{equation}
    \label{LR_alpha_D}
    \Delta_{i,j} \propto \frac{1}{|i-j|_p^{\alpha}} ,
\end{equation}
both with short-range and long-range hopping $t_{i,j}$. The full Hamiltonian,
being quadratic, can be readily diagonalized \cite{Blaizot}, and we denote
again by $\epsilon_k$ the corresponding eigenvalues. The EE of subsystems
can be calculated from the matrices $t_{i,j}$ and $\Delta_{i,j}$, extending
the result (\ref{res}) valid when $\Delta_{i,j}=0$. In the presence of pairings,
indeed, the von Neumann entropy between two subsystems $A$ and $B$,
required to calculate the MI, has to be derived in a different way
with respect to the case $\Delta_{i,j}=0$ when one uses Eq. (\ref{res}).
With $\Delta_{i,j} \neq 0$
the correlations $\langle c_j c_l \rangle$ and
$\langle c_j^{\dagger} c^{\dagger}_l \rangle$ do not  vanish in general.
We thus need to enlarge the correlation matrix to double its size
in order to account for the additional terms. To do so, one define a
matrix $M_{p,q} \equiv \langle (a_{2 p -1} , a_{2 p}) (a_{2 q -1} ,  a_{2 q})
\rangle$,
where $a_{2 q -1} \equiv c_q + c_q^{\dagger}$ and $a_{2 q} \equiv i \,
(c_q - c_q^{\dagger})$ are Majorana fermions,
and the indices $p , q = 1, \dots , l$  run over the $l$ sites of subsystem $A$.
The matrix ${\bf M}$ has $l$ pairs of eigenvalues $1 \pm v_n$,
in terms of which the EE is straightforwardly expressed as in (\ref{res}).
We refer to
\cite{peschel2011} for a presentation of the calculation of the EE
for a generic quadratic form of fermions and for more references on the
subject. 

\section{Short-range models}
\label{sec:sr}

In this Section we consider short-range models, with nearest-neghbor
hopping and $\Delta_{i,j}=0$ (Section \ref{subsec:tb}) or
$\Delta_{i,j}$ nearest-neighbor as well (Section \ref{subsec:kit}).

\subsection{Tight-binding chain}
\label{subsec:tb}
The usual tight-binding chain Hamiltonian reads:
\begin{equation}\label{eqn:Hamiltonian_SR}
 H = -t \sum_{i=1}^{N_S} c_{i}^{\dagger}c_{i+1} + \mathrm{H. c.} \,,
\end{equation}
with the PBC (\ref{eqn:PBC}), so that $c_{N_S+1} = c_1$.
We will consider {\color{black}its generalization in eq.~\eqref{eqn:Hamiltonian}, with a} hopping term of the form:
\begin{equation}
    \label{LR_alpha_real}
    t_{i,j} = \frac{t}{|i-j|_p^{\alpha}} \,,
\end{equation}
the nearest-neighbor case (\ref{eqn:Hamiltonian_SR}) being
obtained {\color{black}in the limit} $\alpha \to \infty$.

The reason for considering (\ref{LR_alpha_real}) is two-fold:
{\it i)} the short-range behaviour is in general expected
to be retrieved for large, finite $\alpha$; {\it ii)}
as discussed in \cite{gori15}, the EE of a subsystem does not depend
on the exponent $\alpha$. Indeed, the correlation matrix 
(\ref{correl_intr}) can be written
as $C_{i,j}=\sum_{l=1}^{N_T} \psi_{l}^{*}(i) \psi_{l}(j)$, where
the $\psi_{\alpha}$ are the eigenfunctions of the matrix $t_{i,j}$.
In the considered translationally invariant case, the quantum number $l$
becomes the wave vector $k$ and the eigenfunctions $\psi_l$ simply plane
waves. So, if the dispersion relation $\epsilon_k$ has a purely monotonous behavior for $k$ either positive or negative,
then the correlation matrix is the same (since one has to sum on the {\it
  same} eigenfunctions). Thus, also the eigenvalues of the correlation
matrix $C_{i,j}$ are equal and so are the EE and, therefore, the MI as well.

\begin{figure}[h]
	\includegraphics[angle=0,width=\columnwidth]{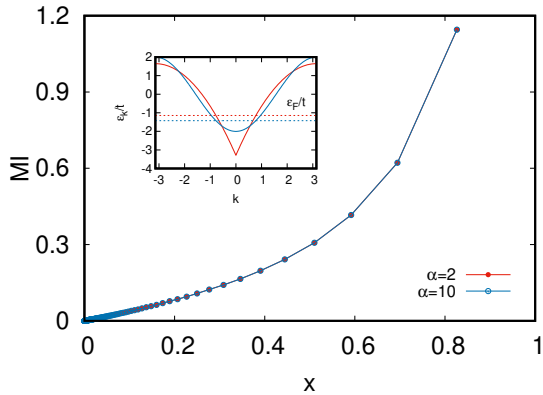}
	\caption{Plot of the MI as a function of the
		conformal four-point ratio (\ref{eqn:cfpr})
		for the hopping $t_{i,j}=t/|i-j|_p^{\alpha}$ in the Hamiltonian
		(\ref{eqn:Hamiltonian}) and
		two different values of $\alpha$: $\alpha=2$ and
		$\alpha=10$. As discussed in the text, checked numerically and seen in the
		figure, the MI does not depend on $\alpha$, and it is the same
		for the short-range limit (\ref{eqn:Hamiltonian_SR}).
		The plot refers to a chain of length $N_S=2004$, with PBC and
		filling fraction $f=0.25$. The four-point ratio is
		calculated varying the distance between
		two subsystems of the same length  $l=10$. Inset:
		the two corresponding dispersion relations $\epsilon_k$ (in units of $t$)
		for $\alpha=2$ and
		$\alpha=10$, depicted together with their Fermi level.
		One sees that, even if the Fermi energies are different,
		the filled states are the same and, as a consequence, the MI is
		the same.} 
	\label{fig:indepalpha}
\end{figure}

\begin{figure}[h]
\dimen0=\textwidth
\advance\dimen0 by -\columnsep
\divide\dimen0 by 2
\noindent\begin{minipage}[t]{\dimen0}
\includegraphics[angle=0,width=\columnwidth]{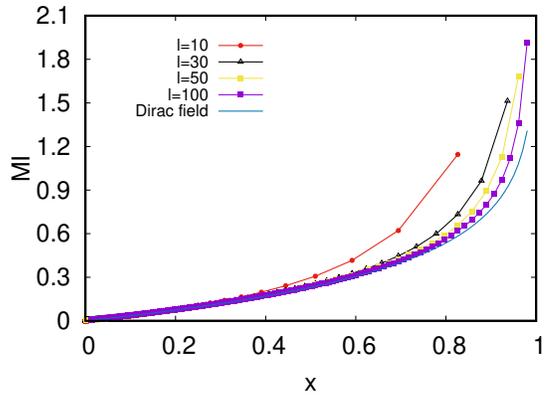}
\end{minipage}	
\hfill
\noindent\begin{minipage}[t]{\dimen0}
\includegraphics[angle=0,width=\columnwidth]{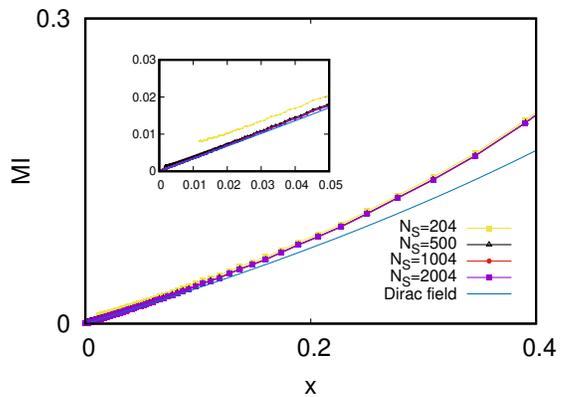}
\end{minipage}
\caption{Top panel: MI for the chain (\ref{eqn:Hamiltonian}) with
hopping (\ref{LR_alpha_real}) with $\alpha=2$ and $N_S=2004$ for several values of the subsystem size $l$. Bottom panel: Scaling of MI with the number of sites   $N_S$ for $l=10$. {\color{black}The blue continuous line represent the analytical prediction in eq.~\eqref{DiracMI}.}
\label{fig:LRPW}}
\end{figure}

Since for any $\alpha$ in (\ref{LR_alpha_real}) the dispersion relation is
indeed monotonous for $k$ either positive or negative, as seen in the inset
of Fig. \ref{fig:indepalpha}, the MI does {\it not} depend on
$\alpha$. The main findings, valid therefore for any positive $\alpha$ in
(\ref{LR_alpha_real}) and in particular for the short-range hopping model
(\ref{eqn:Hamiltonian_SR}), are shown in Fig. \ref{fig:indepalpha} and
Fig. \ref{fig:LRPW}. In Fig. \ref{fig:indepalpha} we plot the MI as a function
of $x$. We see that the MI has a linear behavior
for small $x$ and it is a {\color{black}monotonously increasing} function of $x$.

In Fig. \ref{fig:LRPW} the MI is reported for different values
of the subsystems sizes $l$ at fixed $N_S$ and the same
behavior appears for the considered cases. As supported also
from simulations where the total length of the chain is varied,
it is seen that the curves tends to converge to a monotonic function.

These results can be compared  with the analytical values of the EE for
disjoint subsystems
of Dirac fermions, see
\cite{casini2009} and refs. therein.
This is done again in Fig. \ref{fig:LRPW}:
the analytical values, given by
 \beq
I = \frac{1}{3} \mathrm{log} \, \Big(\frac{1}{1-x} \Big), 
\label{DiracMI}
 \eeq
 are represented by the blue continuous line.
 We find a strongly improving agreement
 between numerics and analytics  as $l$ increases.
 Moreover, for fixed $l$, a very good stability of the data
is achieved for $N_s \sim 1000$.
 
Further analytical results would be desirable to determine
the MI as a function
of $x$ for intermediate and large values of $N_S, l$, and $d$ at
fixed filling to discuss
finite-size corrections, however from the
results presented here it emerges that $x$ is the good quantity to use,
the MI being a monotonic function of $x$. The prediction
(\ref{formula}) is clearly not verified, as expected.

\subsection{Kitaev chain}
\label{subsec:kit}
We consider in this Section the Kitaev Hamiltonian 
\begin{equation}
H = - t \sum_{i=1}^{N_S} \left(c^\dagger_i c_{i+1} + \mathrm{H.c.}\right) 
+ \frac{\Delta}{2} \sum_{i,j=1}^{N_S} \frac{1}{|i-j|_p^{\alpha}} 
\left( c^\dagger_{j} c^\dagger_{i}+c_i c_{j} \right) \, .
\label{Ham}
\end{equation}
focusing on the limit $\alpha \to \infty$. In this limit, one recovers the short-range Kitaev chain \cite{kitaev}, that in
turn can be mapped via Jordan-Wigner
transformations to the 
Ising model in transverse field \cite{libro_cha}.

\begin{figure}[H]
	\dimen0=\textwidth
	\advance\dimen0 by -\columnsep
	\divide\dimen0 by 2
	\noindent\begin{minipage}[t]{\dimen0}
		\includegraphics[angle=0,width=\columnwidth]{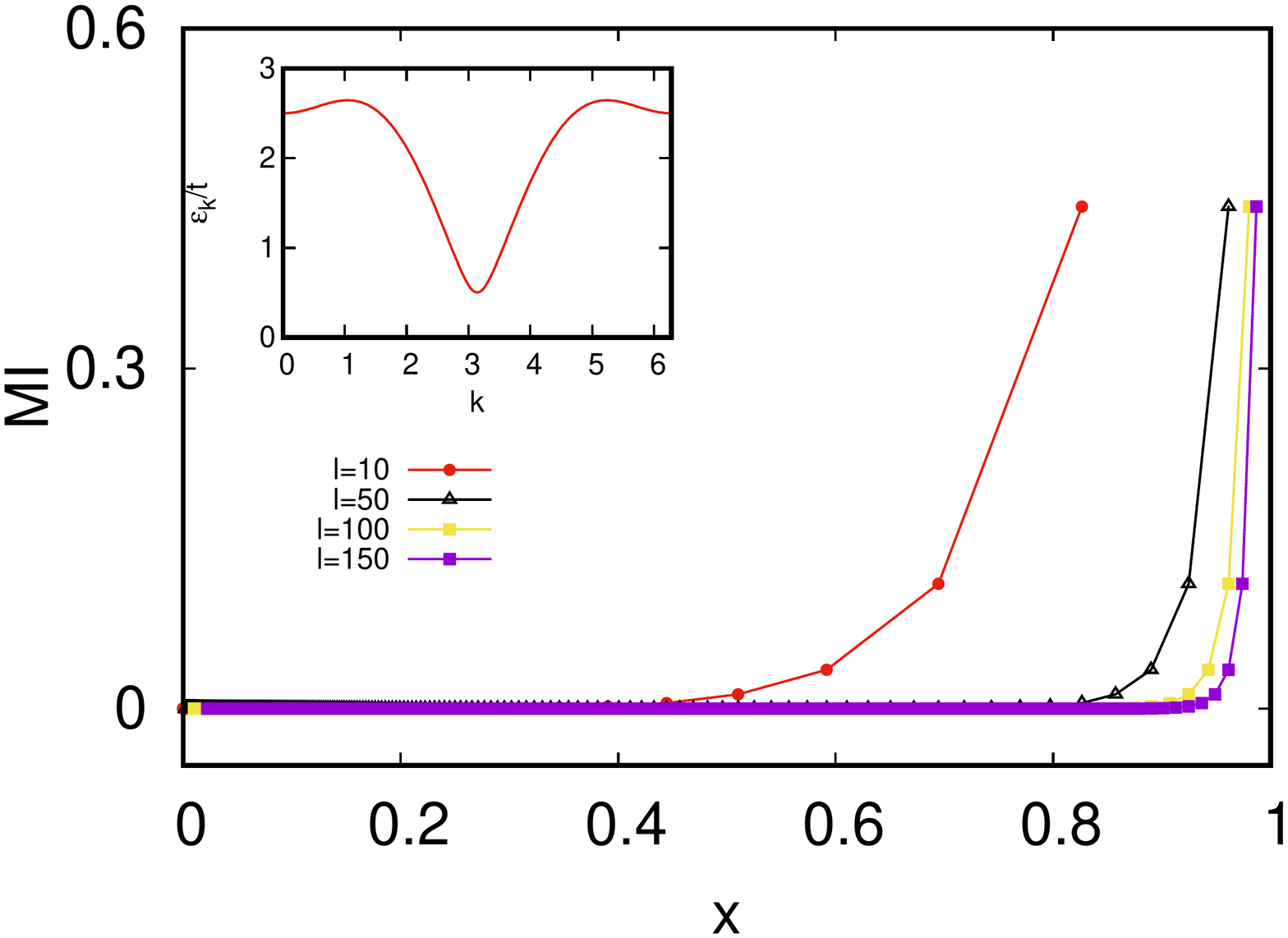}
	\end{minipage}
	\hfill
	\begin{minipage}[t]{\dimen0}
		\includegraphics[angle=0,width=\columnwidth]{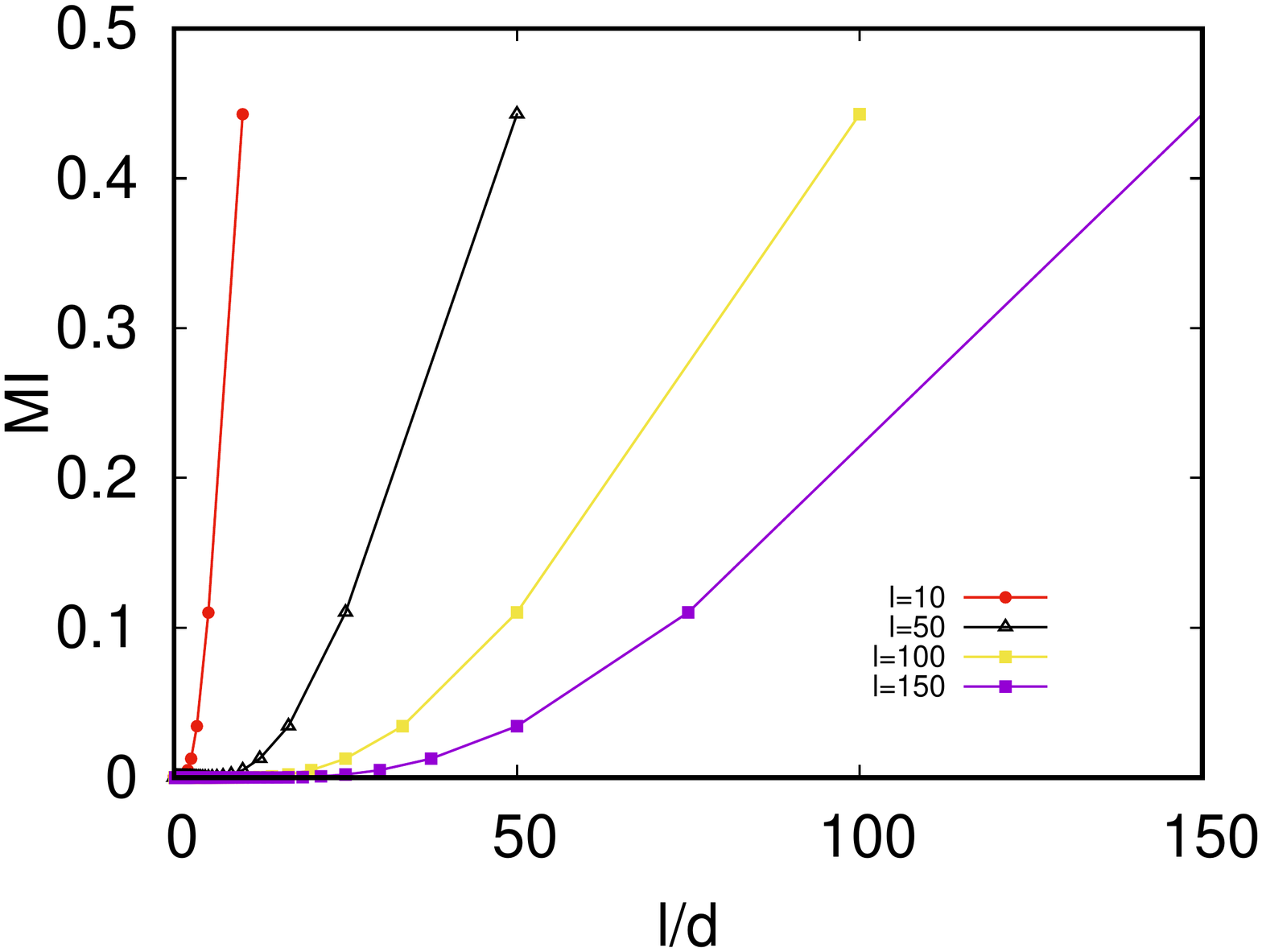}
	\end{minipage}
	\begin{minipage}[t]{\dimen0}
		\includegraphics[angle=0,width=\columnwidth]{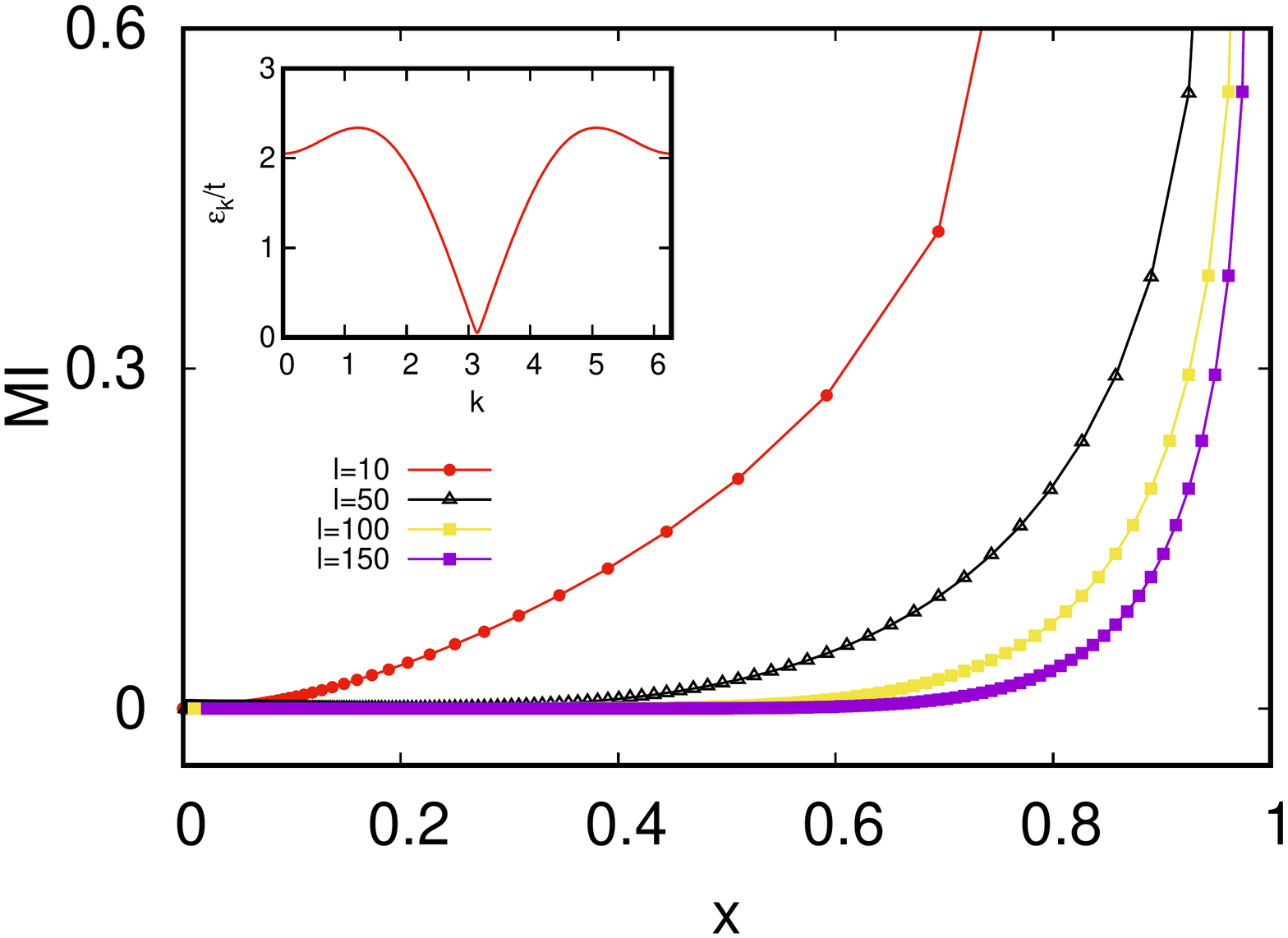}
    \end{minipage}
	\caption{(Upper panel)
		Bogoliubov quasiparticle spectrum (inset) and
		MI as a function of the four-point ratio $x$ 
		for the Kitaev chain (\ref{Ham}) with $\alpha = 1000$,
		using the parameters $N_S = 2004$ and
		$\mu = 1.5$ (corresponding to the
		filling $f = 0.5567$). (Middle panel) 
		Same, but now as a function of $l/d$. {\color{black}(Lower panel) Same as the upper panel, but for $\mu = 1.05$.}}
	\label{fp1}
\end{figure}

For the model (\ref{Ham}), it is convenient to work with anti-periodic boundary
conditions \cite{vodola14}. To fix the {\color{black} average} number of
particles, we introduce a chemical potential by adding the term
$-\mu \sum_{j=1}^{N_S} \left(c_j^\dagger c_j - \frac{1}{2}\right)$
to the Hamiltonian (\ref{Ham}). {\color{black}To compare with previous results, without loss
of generality, we will set $\Delta=2 t=1$, since different choices can be absorbed in a redefinition of $\mu$, therefore measuring the energies in units of $2t$ \cite{vodola14,vodola2016}.}

Presenting the results also for finite $\alpha$,
the spectrum of excitations
is obtained via a Bogoliubov transformation and it results in:
\begin{equation}
\epsilon_k = \sqrt{\left(\mu - \cos{k} \right)^2 + f_{\alpha}^2(k + \pi)} \, ,
\label{eigenv}
\end{equation}
where $k=- \pi + \frac{2\pi}{N_S} \left(n +  \frac{1}{2} \right)$
($0 \leq n< N_S$) and
\begin{equation}
  f_{\alpha} (k) \equiv \sum_{m=1}^{N_S-1} \frac{\sin(m k)}{|m|_p^\alpha} \, . 
\end{equation}
The functions $f_\alpha(k)$ become, in the thermodynamic limit
\cite{vodola2016,lepori16}, a polylogarithmic function \cite{grad,abr,nist}. The ground state of Eq. (\ref{Ham}) is given by 
\begin{equation}
  \ket{\Omega}  =\prod_{n=0}^{N_S/2-1} 
  \left(\cos\theta_{k_n} - i \sin\theta_{k_n} \, {\color{black} c^\dagger_{k_n} 
  c^\dagger_{-k_n}} \right) |0\rangle \ ,
\end{equation}
where $\tan(2\theta_{k_n}) = -\frac{f_{\alpha}(k_n+ \pi)}{\mu -\cos{k_n}}$ and 
it is even under the $Z_2$ fermionic number parity of the
Hamiltonian \cite{muss,fendley2012}.

The spectrum in (\ref{eigenv}) displays a critical line at $\mu = 1$ 
for {\it every} $\alpha$ and a critical semi-line $\mu = -1$ for $\alpha >1$
(and therefore in the short-range Kitaev model $\alpha \to \infty$ as well).

We report in Fig. \ref{fp1}
the MI as a function of the conformal four-point ratio $x$ for
$\alpha =1000$, practically indistinguishable from the short-range Kitaev
model at $\alpha \to \infty$, $N_S= 1000$ and $\mu = 1.5$. In the inset of Fig. \ref{fp1}, we
plot the corresponding spectrum $\epsilon_k$ for the Bogoliubov
quasiparticles. {\color{black} In Fig. \ref{fp1} we plot also MI as a function of $l/d$ for different values of $l$, and the comparison of the two panels shows
  that MI tends to be vanishing for small values of $x$ or $l/d$,
  or better up to certain value of them, and that collapse of data is not observed.} In the considered ranges for the Hamiltonian parameters, the model satisfies an area law for the EE, and correspondingly 
\textcolor{black}{the MI appears to vanish for $x<1/2$}.
However, the region
in which MI vanished increases beyond $1/2$ approaching the
thermodynamic limit.

Finally, studying the MI on the massless line $\mu = 1$,
one finds a monotonic growth of the MI with $x$,
qualitatively equal to those in Figs. \ref{fig:indepalpha}
and \ref{fig:LRPW}, and
corresponding to a logarithmic violation of the area law by the von
Neumann entropy. 
For completeness, in Fig. \ref{fp1} we also report the MI
close to the critical line (choosing $\mu=1.05$). There, a comparison
with the predictions of the Dirac theory \cite{casini2009}, similarly as in the previous 
Section, would require a huge fine tuning of the Dirac mass.

\section{Long-range hopping} 
\label{sec:tb}
In this Section we consider {\color{black}several} models
with long-range hopping. 

For the model (\ref{eqn:Hamiltonian}), without pairing terms ($\Delta_{i,j}=0$)
and long-range hopping (\ref{LR_alpha_real}) given by
$t_{i,j}=t/|i-j|_p^{\alpha}$, one can readily determine the
energy spectrum as 
\begin{equation}
\epsilon_k = - 2t \, \ell_{\alpha}(k) \,,
\label{disprel}
\end{equation}
where $k=2\pi n_k/N_S$ belongs to the first Brillouin zone,
$n_k=-N_S/2,\cdots,N_S/2-1$, and 
\begin{equation}
\ell_\alpha(k)= \sum_{n=1}^{\infty} \frac{\cos{(n k)}}{n^\alpha} \,.
\label{elle_funct_app}
\end{equation}

As commented in Section \ref{subsec:tb}, the dispersion relation
(\ref{disprel}) is monotonous in $k$ as in the short-range (see the inset
of Fig. \ref{fig:indepalpha}).
The long-range hopping (\ref{LR_alpha_real}) only changes the
values of the single-particle energies, and the MI is exactly the same
as in the short-range model, as discussed in Section \ref{subsec:tb} and
illustrated in Figs. \ref{fig:indepalpha} and \ref{fig:LRPW}.
We conclude that long-rangedness is not enough to change the
MI properties: one also needs (if pairing terms are absent)
to change the behavior of the
single-particle spectrum, and, in particular, the Fermi surface.
This is discussed in the next Section.

\subsection{Fractal Fermi surface}
\label{sec:fractal}
\begin{figure}[h]
	\includegraphics[angle=0,width=\columnwidth]{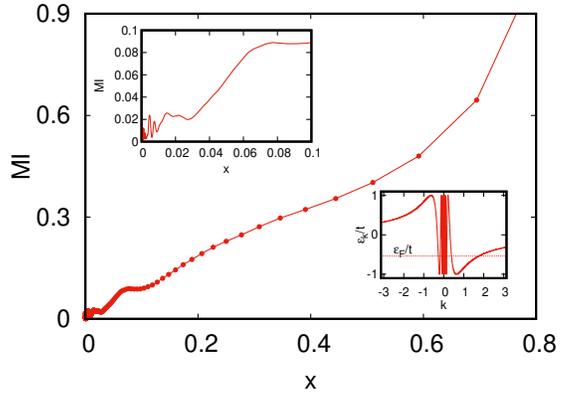}
	\caption{Plot of the MI as a function of the conformal four-point ratio $x$
		for the chain with dispersion (\ref{sin_1_k}) and
		$\gamma=1$, $f=0.25$ and $N_S=1004$. The four-point ratio
		is calculated varying the distance between two subsystems of the same
		length $l=10$. In the top-left inset the small-$x$ region is enlarged.
		In the bottom-right inset dispersion is plotted with the
		Fermi level (dotted line).} 
	\label{fig:frac01}
\end{figure}

We assume here that 
the Hamiltonian is such that the single-particle energy spectrum has the form 
\begin{equation}
\varepsilon_k=-t \sin{\left( \frac{1}{k^\gamma} \right)},
\label{sin_1_k}
\end{equation}
where $\gamma$ is a positive odd
integer. {\color{black} Hamiltonian \eqref{sin_1_k} is written
  in momentum space, and when written in real space the hopping $t_{ij}$ assumes
  a complicated and oscillating in sign form, having as a first approximation
  a power-law envelope.} As shown in \cite{gori15},
at certain fillings, the Fermi surface of this model has a fractal topology.
For instance, at half-filling $f=0.5$,
the Fermi energy is zero, 
with the point $k=0$ being an accumulation point, see {\color{black} the bottom right inset of}
Fig. \ref{fig:frac01} .
The Fermi surface has 
box counting dimension $d_{box}=\frac{\gamma}{\gamma+1}$, so that $d_{box}=1/2$ for $\gamma=1$. From finite size numerical data one sees
that the EE violates the area law as $S_A \sim L^\beta$,
where $\beta = d_{box}$ \cite{gori15}.

\begin{figure}[h]
	\includegraphics[angle=0,width=\columnwidth]{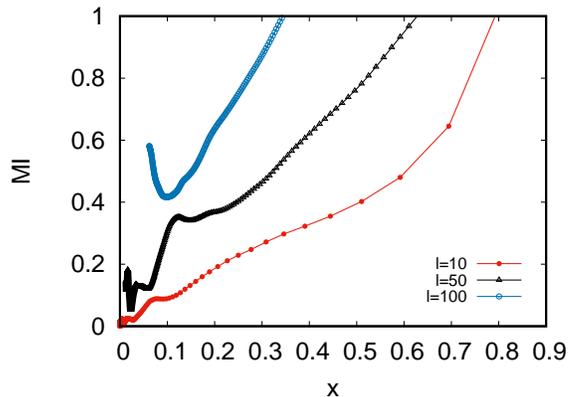}
	\caption{MI  vs $x$ for the chain with dispersion (\ref{sin_1_k}) and
		$\gamma=1$, $f=0.25$ and $N_S=1004$. The four-point ratio
		is calculated varying the distance between two subsystems of the same
		length $l=10,20,50$.
	} 
	\label{fig:frac02}
\end{figure}

\begin{figure}[t]
\includegraphics[angle=0,width=\columnwidth]{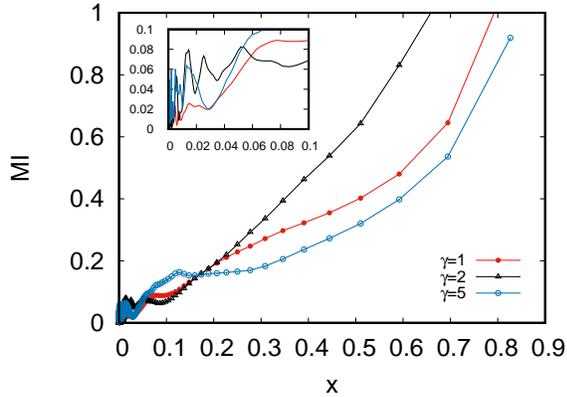}
\caption{
MI {\it vs} $x$ for the  chain with dispersion (\ref{sin_1_k})
  and $f=0.25$, $N_S=1004$ for different values of  $\gamma$.
  The four-point ratio is calculated varying the distance between
  two subsystems of the same length $l=10$.
  In the top-left inset the small-$x$ region is enlarged.} 
\label{fig:frac03}
\end{figure}

Our results for the MI are reported in Figs. \ref{fig:frac01}-\ref{fig:frac03}. 
The observed behavior differs from the previous models considered and
displays very peculiar and unique features at small $x$.
In Fig. \ref{fig:frac01} the MI as a function of $x$ is shown at a fixed
value of $\gamma$. It is seen that the non-logarithmic violation
of the area law makes the MI non-monotonous  at small $x$.
In Fig. \ref{fig:frac02} we plot the results obtained for different sizes
of the subsystem size $l$: 
the MI does not seem to be dependent solely on the conformal
four-point ratio $x$, since, in this model, varying the subsystems'
length produces different outcomes. Moreover, a non-logarithmic
violation of the area law violates the basic assumption of holography and
thus we cannot expect Eq. (\ref{formula}) to apply here, since both conformal invariance and AdS/CFT duality are not respected. Our numerical
results support this conclusion. In Fig. \ref{fig:frac03},
the MI is plotted for different values of $\gamma$, {\color{black}and thus of exponent $\beta$, defined after Eq. \eqref{sin_1_k}, which quantifies} the deviation from the area law. {\color{black} We considered other combinations
  of system parameters different from $x$ such as $l/d$, but we found
  that they did not introduce a simplification of the obtained results and} {\color{black}thus we do not presents plots of such results.}
It is found that the larger is the deviation from the area law, the more
non-monotonic the MI is. 
{\color{black}Figure \ref{fig:frac03}} also indicates that the Bell-pairs responsible for the EE have a specific distribution in space also at very large distances.
{\color{black}Although an analytical determination of the peaks position in Figs. \ref{fig:frac01}-\ref{fig:frac02} is a very non-trivial task, we can qualitatively comment on their difference compared to the long-range model with a phase which we study in the following Section \ref{sec:LRPLPhase}. In the latter case, the EE follows a volume law, while here the deviation from the area law is weaker and the fractal Fermi surfaces selects values of the momentum $k$ that do not necessarily alternate \cite{gori15}. Thus, the Bell pairs form not only close to antipodal sites but also between closer sites resulting in the complicated peak patterns observed in Figs. \ref{fig:frac01}-\ref{fig:frac02}.}

\subsection{Other models with non-logarithmic violations of the area law}
\label{subsec:volume}
Other models exhibiting non-logarithmic violations of the area law
are considered in the following Sections \ref{sec:LRPLPhase}
and \ref{sec:selhop}. In both cases the pairing terms are absent, their effect
being studied in Section \ref{sec:pairing}.

\subsubsection{Long-range power-law model with a phase}
\label{sec:LRPLPhase}

A possible way to modify the structure of the Fermi surface, is to
introduce a space-dependent phase in the hopping matrix 
\begin{equation}\label{magnetic_phase}
t_{i,j} = \left\{
\begin{array}{ll}
0 \, , & i=j,\\
\frac{t e^{i \phi d (i-j)} }{|i-j|_{p}^\alpha} \, , & i \neq j \, ,
\end{array} \right.
\end{equation}
where $\phi = \frac{2 \pi}{N_S} \Phi$ (with $\Phi$ a constant) and
$d(m)$ is the oriented distance, defined as
\begin{equation}
   d(m) \equiv  \left\{
   \begin{array}{ll}
   m & |m| \le N_S - |m|,\\
   {}\\
   -N_S + |m| & {\rm otherwise} \, .
   \end{array} \right.
\end{equation}
The energy spectrum is found to be
$\epsilon_k = - 2t\ell_{\alpha}(k)$ with  
\begin{equation}
\ell_\alpha(k, \phi;N_S) = \sum_{n=1}^{N_S/2} \frac{\cos{[n(k + \phi)]}}{n^\alpha} \, .
\label{elle_funct_app_phi}
\end{equation}

\begin{figure}[t]
	\includegraphics[angle=0,width=\columnwidth]{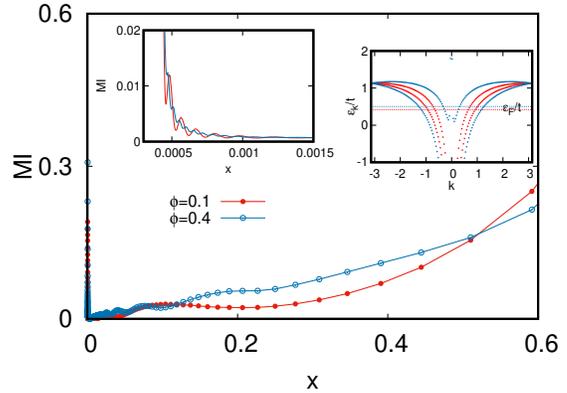}
	\caption{MI for the long-range model (\ref{eqn:Hamiltonian}) with the phase-modulated hopping (\ref{magnetic_phase}) with two values of the phase $\phi=0.1,
		0.4$ and parameters $N_S=1004$, $\alpha=0.3$, $l=10$, and $f=0.25$.
		Left inset: MI behaviour for small $x$. Right inset: energy spectrum for
		the two values of $\phi$. 
	}
	\label{fig:LRphase}
\end{figure}

\begin{figure}[t]
\includegraphics[angle=0,width=\columnwidth]{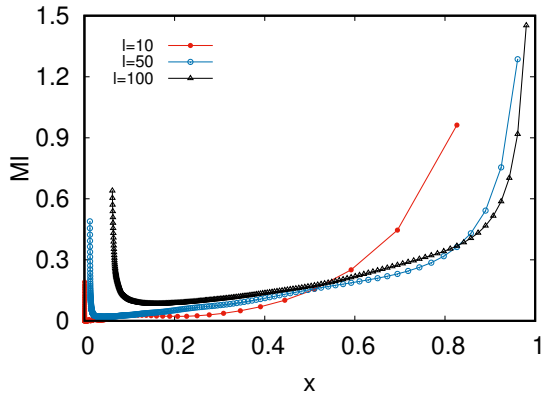}
\caption{MI for the long-range model (\ref{eqn:Hamiltonian})
  with the phase-modulated hopping (\ref{magnetic_phase}) for
  $N_S=1004$, $\alpha=0.3$, $f=0.25$, $\phi=0.1$ and different values of $l$.
}
\label{fig:LRphase01}
\end{figure}

While for $\phi=0$ this function is monotonic for $k \in [0,\pi]$,
for non-zero $\phi$ there is a critical value of $\alpha_c (\phi)<1$ such
that for $\alpha < \alpha_c (\phi) $, at half filling $f=1/2$, the
energies (\ref{elle_funct_app_phi}) are rapidly oscillating and the momenta
$k$ are occupied in an alternating way between even and odd quantum numbers,
as depicted in the right panel of Fig. \ref{fig:LRphase}. As a consequence,
a Bell-paired-like ground state is obtained, with a volume law EE \cite{gori15}.
We notice that for $\alpha \le 1$ the ground-state energy in the
thermodynamic limit diverges. Nonetheless, one can make the energy
extensive by the so-called Kac rescaling \cite{libro}.  
As reported in Fig. \ref{fig:LRphase}, the MI shows a non-monotonic
behavior with $x$, emerging at small $x$ (see left inset),
where peaks develop. One also sees a clear peak at $x \to 0$.
We attribute the presence of such non-monotonicity and peaks
to the formation of Bell pairs,
which are the reason for the (non-logarithmic) deviation
from the area law. This will be more evident
in the models studied in the
following Section \ref{sec:selhop}. {\color{black} Notice that by plotting MI as a
  function of $l/d$ we obtain similar results.}
Also for the long-range power-law model with a phase,
as for the model of Section
\ref{sec:fractal},
the four-point ratio $x$ is found to be not
sufficient to capture the whole structure of the MI, since as illustrated
in Fig. \ref{fig:LRphase01}, results with different values of $l$ but the same
$x$ exhibit different results for the MI. However,
non-monotonous behaviour and the increase for $x\to 0$ are seen.

\subsubsection{Selective hopping}
\label{sec:selhop}
We study in this Section the "antipodal" model in which each-site
is connected via a hopping term to the more distant site
in the chain (at the antipodes). This model
has a maximally entangled ground-state with the maximum
number of Bell pairs, and it exhibits perfect, {\color{black}maximal} volume law \cite{gori15}. 
{\color{black} As such, as we will shortly show, the MI vanishes identically, due to the monogamy of entanglement.}
To understand the role of deviations from {\color{black}a maximal} volume law, we
consider a chain with supplementary selective
long-range hoppings, where
the particle can only hop between sites centered around two distances,
denoted by $s_{1,2}$, in a window of $2r+1$ sites
($r$ being an integer).
The Hamiltonian reads 
\begin{equation}
H=-\sum_{j=1}^{N_S}\sum_{q=-r}^{r}\left(t_{1}c_{j}^{\dagger}c_{j+s_{1}+q}+t_{2}c_{j}^{\dagger}c_{j+s_{2}+q}+ \mathrm{H.c.}\right) \, ,
\label{selhopHam}
\end{equation}
where PBC are implied. This model is readily diagonalized in Fourier space, with
energies
\begin{equation}
\epsilon_{k}=
-2\Big[t_{1}\cos\left( k s_{1} \right)
+t_{2}\cos\left( k s_{2} \right)\Big] \cos\left( k r \right)
\left[1+\frac{\tan\left( k r \right)}{\tan\left(k/2\right)}\right] \ .
\end{equation}
We now look at the MI for three specific cases: 

\begin{figure}[h]
\includegraphics[angle=0,width=\columnwidth]{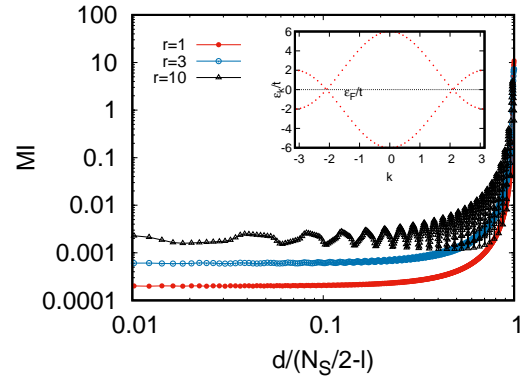}
\includegraphics[angle=0,width=\columnwidth]{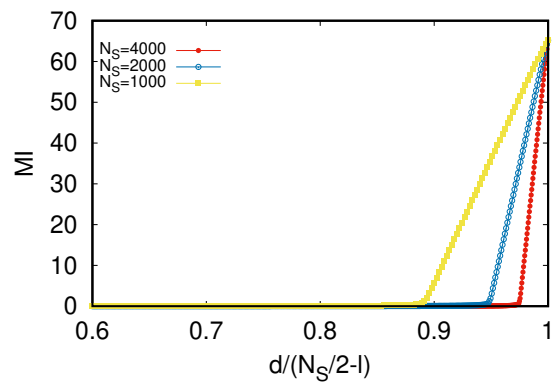}
\caption{MI for the antipodal hopping model (\ref{selhopHam}),
  with $s_1=N_S/2$ and $s_2=0$. Top panel: MI for $N_S=1000$ and $r=1,\,3,\, 10$.  Inset: energy spectrum for $r=1$. 
  Bottom panel: MI for $N_S=1000,\, 2000,\, 4000$ and $r=1$.}
\label{fig:Antinodal}
\end{figure}

{\em Antipodal Hopping: $s_1=N_S/2$, $t_2=0$ ($t_1=t$):}
Let us first set $r=0$ (pure antipodal hopping).
This means that we allow particles to hop only
between the antipodal points and back
(again, we take $N_S$ to be an even number).
The single-particle spectrum reduces to
\begin{equation}
\epsilon_{k}= -2 t \cos\left( \frac{k N_S}{2} \right)
= -2t\left(-1\right)^{n_k} \ , \quad k = \frac{2 \pi}{N_S} n_k \ ,
\end{equation}
which shows that there are two branches
in the flat dispersion, with positive (negative) energy for odd (even)
wave numbers.
At half-filling, we thus have an alternating Fermi surface and {\color{black}an EE
ruled by a perfect, maximal volume law.} In fact, this choice of parameters
creates a long-range Bell-paired state of maximal EE, since each particle
is delocalized between two sites half-system apart.
Remarkably, we 
find that such high entanglement content does not translate into the MI,
which is in fact vanishing for any {\color{black} distance}. This can be explained
as a reflection of the entanglement monogamy
\cite{coffman2000,osborne2006} of the model.
Thus, one needs systems with a lower {\color{black}bipartite} EE.

Allowing for $r \ne 0$ reduces the EE. For $r=1$ one has
\begin{equation}
\epsilon_{k}=-2t\left(-1\right)^{n_k}\left[1+2\cos\left(\frac{2\pi}{N_S} n_k\right)\right] \ .
\end{equation}
Fig. \ref{fig:Antinodal} displays the MI for this case. {\color{black} Notice that here, and in the next Figs. \ref{fig:Antinodal-1} and \ref{fig:selec2} we prefer to plot MI not as a function of $x$,} {\color{black}but of a scaled distance $d$ between the intervals.}
We see that the reduction in EE corresponds to a peak of MI
close to {\color{black} $d=\frac{N_S}{2}-l$}, almost vanishing otherwise. 
{\color{black} Oscillations in Fig.9 are really small and  correspond to small variation of the entanglement due to longer hopping paths connecting the two partitions. The length of this path, and so the period $T$ of the oscillations,  decreases with increasing $r$. It appears that  $T\propto 1/r$}

{\em Single hopping with $s_1=s$, $t_2=0$ ($t_1=t$): }
Here we have the dispersion relation 
\begin{equation}
\epsilon_{k}=-2t\cos\frac{2\pi s \: n_k}{N_S} 
\end{equation}
with $r=0$, and 
\begin{equation}
\epsilon_{k}=
- 2 t \cos\left(\frac{2\pi s \: n_k}{N_S} \right)
\left[1+\frac{\tan\left(\frac{2\pi r \: n_k}{N_S} \right)}{\tan\left(\frac{\pi n_k}{N_S}\right)}\right]
\cos\left(\frac{2\pi r \: n_k}{N_S}\right) 
\end{equation}
for general $r$, from which the Fermi surfaces can be worked out.
The EE violates the area law and the MI is plotted
for different values of $s$ in Fig. \ref{fig:Antinodal-1}
for $r=0$. Similar results are found for other values of $r$ 
(not reported). 
One sees that there is a {\color{black} main peak corresponding to the Bell pair at the given $s$ where the maximum amount of Bell pairs is expected.
Other  secondary peaks appear at distances that are a multiple of $s$}.

\begin{figure}[h]
\dimen0=\textwidth
	\advance\dimen0 by -\columnsep
	\divide\dimen0 by 2
	\noindent
	\hfill
    \begin{minipage}[t]{\dimen0}
		\includegraphics[angle=0,width=\columnwidth]{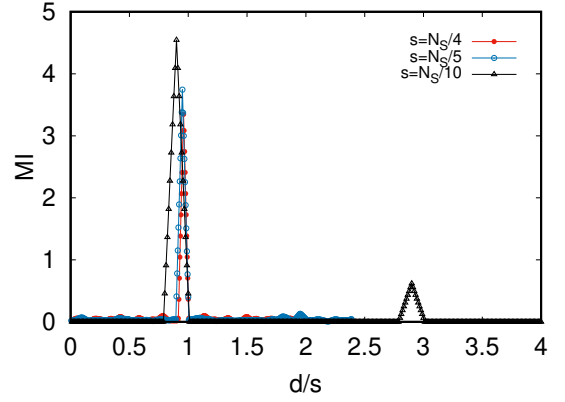}
		\vskip -1cm
		\includegraphics[angle=0,width=\columnwidth]{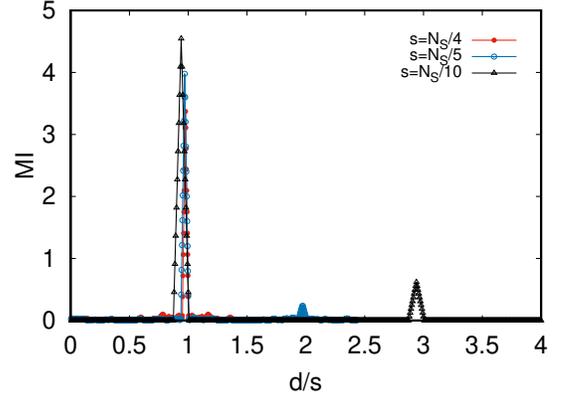}
    \end{minipage}
	\vskip-1cm
	       \caption{MI for the selected hopping model (\ref{selhopHam})
                  with $t_2=0$, $f=0.5$, 
                  $r=0$ and different values $s=N_S/4, N_S/5, N_S/10$.
                  We consider different chain lengths
                  $N_S=900$ (top panel) and $1500$ (bottom panel).}
	\label{fig:Antinodal-1}
\end{figure}

\begin{figure}[h!]
	\includegraphics[angle=0,width=\columnwidth]{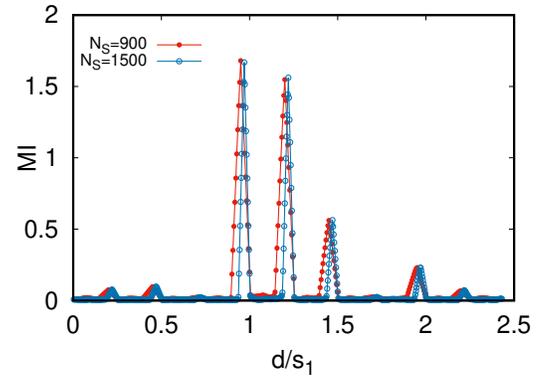}
	\vskip -1cm
	\caption{MI for the selected hopping model (\ref{selhopHam})
		from any site only to two distant sites ($r=0$, $l=10$ and $f=0.5$)  
		$s_1=N_S/4$ and $s_2=N_S/5$. 
	}
	\label{fig:selec2}
\end{figure}

{\em Hopping exactly to two distant sites with $r=0$: }
With hopping allowed only between sites with distances
$s_{1}$ and $s_2$, one has 
\begin{equation}
\epsilon_{k}=-2\left[t_{1}\cos\left(\frac{2 \pi s_1 \: n_k}{N_S} \right)
+t_{2}\cos\left(\frac{2 \pi s_2 \: n_k}{N_S} \right)\right] \ .
\end{equation}
For $s_1 = N_S/2$ and $s_2 = N_S/4$ at half-filling $f=1/2$, the
single-particle spectrum is following a zig-zag behaviour,
with odd wavenumber states filled in the ground state.
As for the case of Fig. \ref{fig:Antinodal},
this is a maximal EE case, where two particles are delocalized over four sites
($N_S/4$ apart), and the MI vanishes. 
In Fig. \ref{fig:selec2}, we show the MI for  $s_1 = N_S/4$ and $s_2 = N_S/5$,
at half-filling $f=1/2$.
The EE is lowered, but
the MI shows peaks for {\color{black} $d=s_1$ , $d=s_2$ and multiple distances related to the corresponding Bell pairs}.

\begin{figure}[t]
	\includegraphics[angle=0,width=\columnwidth]{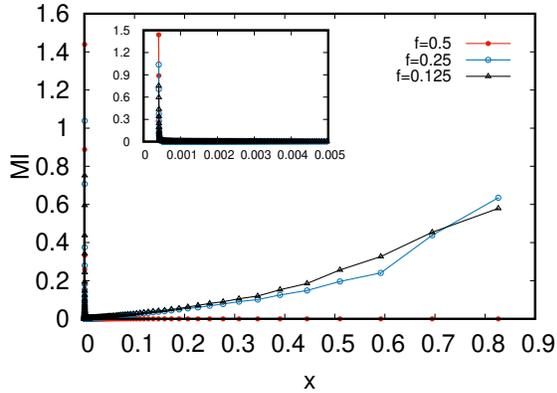}
	\caption{MI for the model (\ref{selhopHam})
                  with hopping from any site only to the two sites
                  surrounding the antipodal point ($s_{1,2}=N_S/2 \pm 1$),
                  with the right panel zooming
                  the small $x$ behavior.}
	\label{fig:LRantinodal}
\end{figure}

We also considered the case $s_1 = N_S/2 -1, s_2 = N_S/2 +1$ ($t_1=t_2=t$)
with $r=0$,
i.e.
\begin{equation}
H=-t\sum_{j}\left(c_{j}^{\dagger}c_{j+\frac{N}{2}+1}+c_{j}^{\dagger}c_{j+\frac{N}{2}-1}+\mathrm{H.c.}\right),
\end{equation}
with single-particle spectrum
\begin{equation}
\epsilon_{k}=-4t\left(-1\right)^{n_k}\cos\frac{2\pi}{N} n_k.
\end{equation}
The MI is depicted in Fig. \ref{fig:LRantinodal} and
shows similarities with the case in Fig. \ref{fig:LRphase},
where one has indeed a volume law, but not the perfect (i.e., maximal)
volume law. 

\section{Long-range Kitaev chains} 
\label{sec:pairing}
We consider in this Section the effect of a long-range counterpart
of the fermionic pairing
as given in Eq.\eqref{eqn:Hamiltonian_P}, both in  the presence
of short-range (Section \ref{sec:LRpairing}) and
long-range (Section \ref{sec:LRHopPairing}) hopping.

\subsection{Short-range hopping and long-range pairing}
\label{sec:LRpairing}
Let us consider here long-range pairing ($\Delta_{i,j}=\frac{\Delta}{2|i-j|_p^\alpha|}$)
and nearest-neighbour hopping ($t_{i,j}=t$ only if $i=j\pm 1$ and zero
otherwise). The MI for the limit $\alpha \to \infty$ has been discussed
in Section \ref{subsec:kit}, where we also
reported the energy spectrum for finite
$\alpha$ in Eq. (\ref{eigenv}). As in Section \ref{subsec:kit} we consider
$\Delta=2t$.

We remind that the spectrum (\ref{eigenv})
displays a critical line at $\mu = 1$ 
for every $\alpha$ and a critical semi-line $\mu = -1$ for $\alpha >1$.  
Moreover, if $\mu \neq -1$ the velocity of quasiparticle with $k = \pm \pi$
diverges if $\alpha \leq \frac{3}{2}$, while it diverges at $\alpha \leq 2$ 
if $\mu = -1$ \cite{vodola14}.
Below $\alpha = 1$ and at every value of $\mu$, new phases occur \cite{vodola14,lepori2016,pezze2017}.
There, the area law for the
EE is logarithmically violated \cite{ares2015,lepori2016}. Moreover the boundary Majorana modes,
present above $\alpha=1$ if $|\mu| <1$, become massive and disappear. 
Notably, this transition to the new phases at $\alpha = 1$ occurs without
any mass gap closure, as a consequence of the large space correlations
induced by the long-range pairing. This transition is signaled by the ground-state fidelity
\cite{pezze2017}.

\begin{figure}[h]
		\includegraphics[width=\columnwidth]{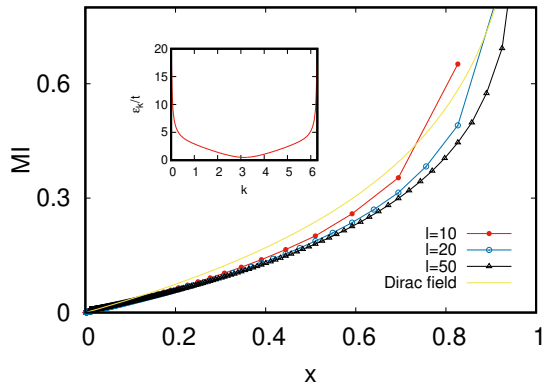}
	        \caption{MI as a
                  function of the four-point ration $x$ for the long-range
                  Kitaev model with $\alpha=0.5$ and short-range hopping,
                  with $N_S = 1000$, $\mu = 1.5$ (corresponding to an average occupation filling $f = 0.5567$). 
                Inset: Bogoliubov quasiparticle spectrum.}
	\label{fp2}
\end{figure}

 \begin{figure}[h!]
	\includegraphics[width=\columnwidth]{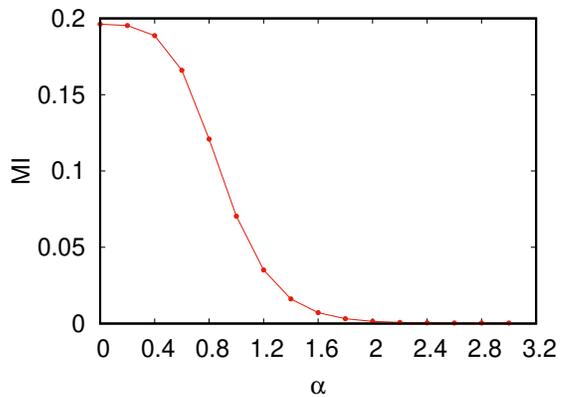}
	\vskip -1cm
	\caption{MI at $x = 0.484$ for the long-range
		Kitaev model {\color{black}as a function of the decay exponent $\alpha$ for the pairing terms} and short-range hopping,
		with $N_S = 200$ and $\mu = 1.5$.}
	\label{MIa1}
\end{figure}

Also for the MI, qualitatively different results are found passing
through the line $\alpha=1$. 
We report in Fig. \ref{fp2} 
the MI, as a function of the conformal four-point ratio $x$
for $\alpha=0.5$, for a closed chain with length $N_S= 1000$ and
$\mu = 1.5$. In the inset, we plot the
Bogoliubov spectrum  $\epsilon_k$.
For $\alpha \gtrsim 1$, the model satisfies an area law for the EE
and the MI is qualitatively similar to that of Fig. \ref{fp1}:
it appears to be vanishing for small $x$.
Instead, for $\alpha \lesssim 1$, the EE is logarithmically violated, and the 
MI increases monotonically with $x$, qualitatively similar to that in Eq. \eqref{fp2}.
This is not due to the fact the model is gapless (see the inset),
but rather to the fact that it has a such long-ranged pairing
that the correlation functions are power-law decaying. This make the MI
of the system similar to the tight-binding model studied in Sections 
\ref{subsec:tb} and \ref{sec:tb}.

Around the line $a = 1$,  the change in behaviour is smooth at finite size $N_s$, similarly to the EE \cite{vodola14}.
This behaviour is shown in Fig. \ref{MIa1}, {\color{black}where MI is reported as a function of $\alpha$ for a value of $x$ close to $1/2$.}
Finally, on the massless line $\mu = 1$, also at $\alpha < 1$,
we find again a monotonic growth of the MI with $x$, similar to
the short-range Kitaev chain.

\begin{figure}[t]
	\dimen0=\textwidth
	\advance\dimen0 by -\columnsep
	\divide\dimen0 by 2
	\noindent\begin{minipage}[t]{\dimen0}
		\includegraphics[width=\columnwidth]{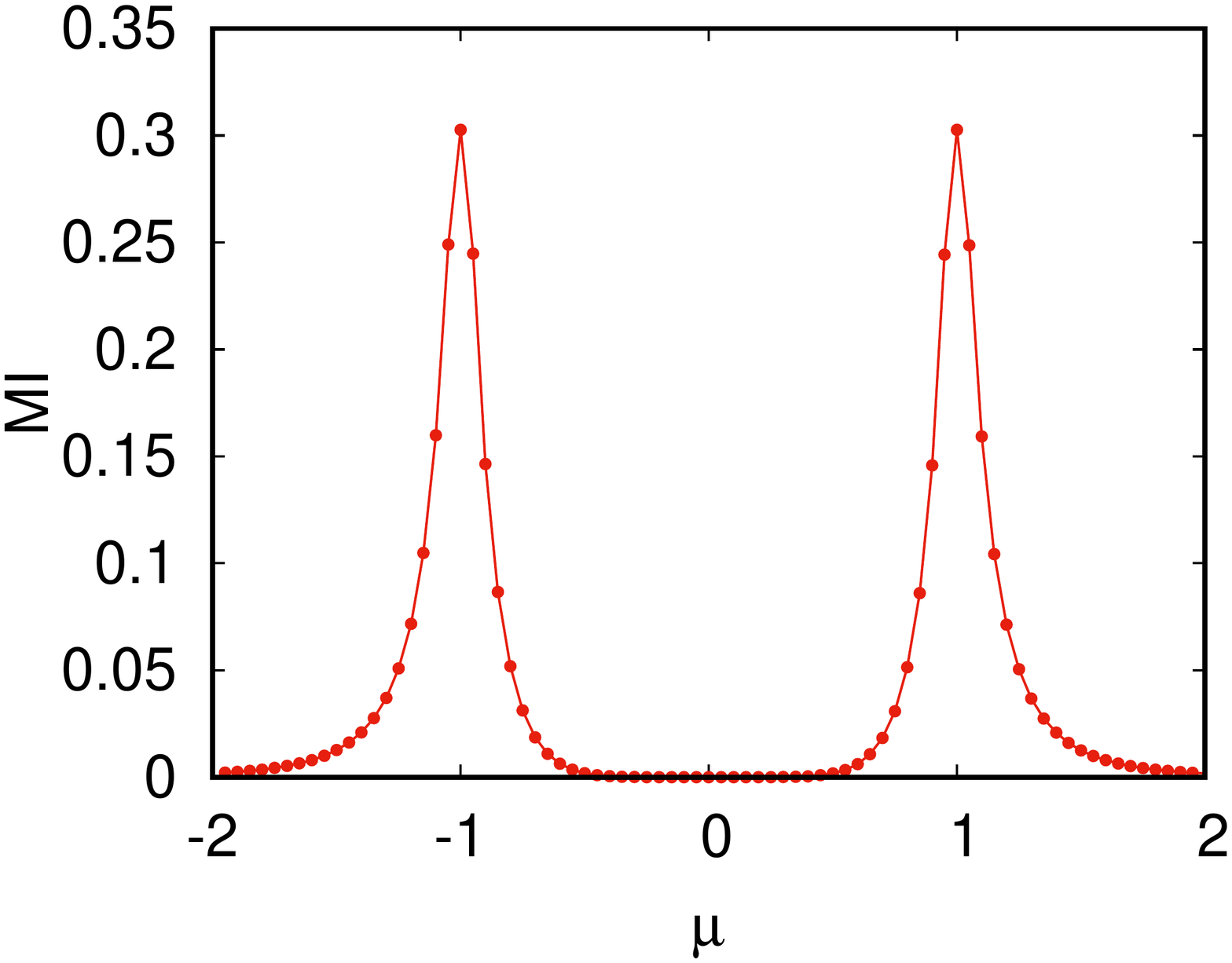}
	\end{minipage}
	\vskip -1cm
	\begin{minipage}[t]{\dimen0}
		\includegraphics[width=\columnwidth]{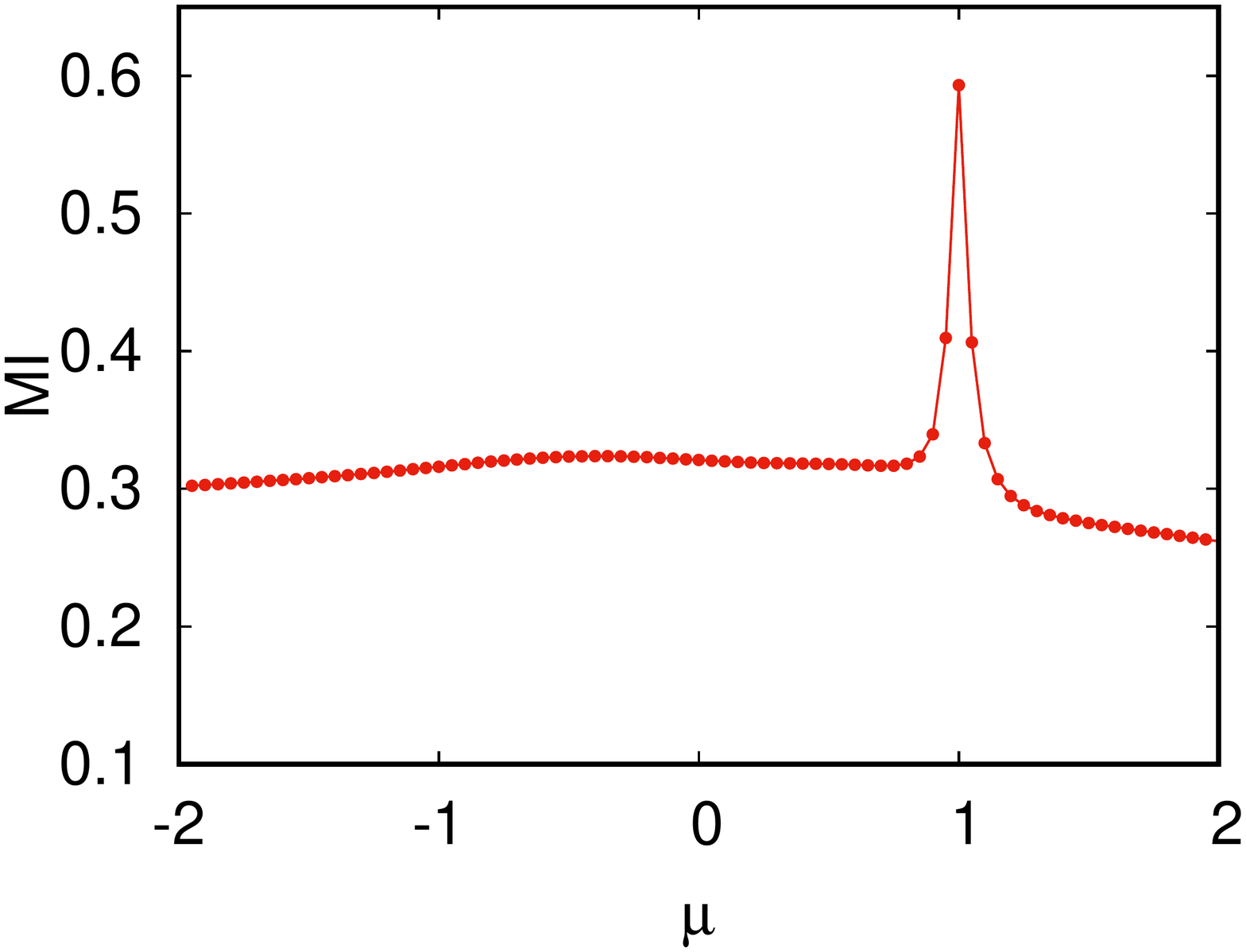}
	\end{minipage}
	\vskip -1cm
	\caption{MI as a function of the chemical potential $\mu$
		for the Kitaev chain with long-range pairing and short-range hopping 
		with  $\alpha = 10$ (top) and $\alpha=0.5$ (bottom) and
		$N_S = 2000$, $l=16$, and $d=4$ sites.
	}
	\label{fp5}
\end{figure}

In Fig. \ref{fp2}, a comparison is made against
the analytical values of the {\color{black}MI, derived from the EE} for disjoint subsystems
of Dirac fermions.
We find that the agreement is rather poor. However, this mismatch is
of course
expected, 
since a Dirac structure does not hold for $\alpha <2$, not even around the massless (semi-)lines at $\mu = \pm 1$, where conformal invariance is broken by the long-range coupling \cite{lepori16}.
We recall that instead the same Dirac structure holds around  $\mu = \pm 1$ and  for $\alpha >2$ \cite{muss}.

\begin{figure}
	\includegraphics[width=\columnwidth]{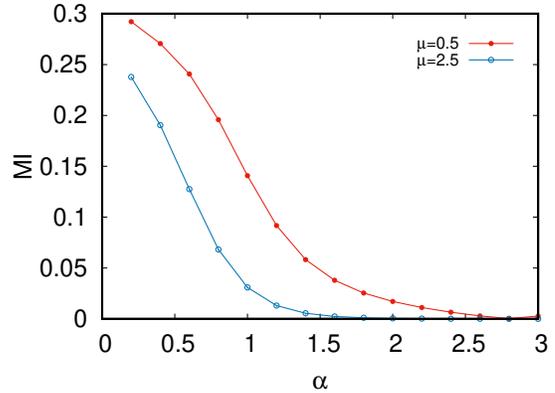}
	\vskip -1cm
	\caption{MI as a function of the power of long-range decay
             $\alpha$ of the pairing for $\mu = 0.5$  and
                  $\mu=2.5$ with $N_S = 1000$, $l=16$, $d=4$ sites.}
	\label{fp6}
\end{figure}

\begin{figure}
	\includegraphics[width=\columnwidth]{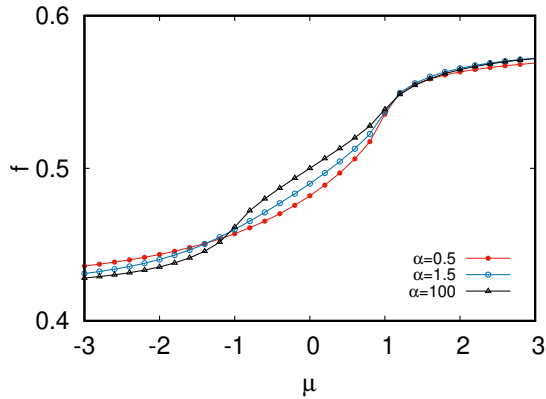}
	\vskip -1cm
	\caption{Filling vs chemical potential for different values
              of $\alpha$.}
	\label{fp6_bis}
\end{figure}

The MI can be used to detect quantum phase transitions. To show this, we
keep the four-point ratio fixed and swipe for the phase-diagram parameters
$\mu$ and $\alpha$. In Fig. \ref{fp5} we show in the top panel the
MI as a function of $\mu$ for a chain with length $N_S = 1000$,
$l= 16$, $d= 4$, and $\alpha = 10$.
Two peaks are observed at $\mu = \pm 1$, in correspondence to the two
critical points. 
Similarly, in the {\color{black}bottom} panel we report the MI vs $\mu$ for $\alpha=0.5$
and the same choices for the other parameters. We
observe a single peak, in correspondence to the unique critical point at
$\mu = 1$. \\
\indent In Fig. \ref{fp6} we plot the MI vs $\alpha$ for $\mu = 0.5$ and
$\mu = 2.5$, and the other parameters as before.
We observe a substantial increase for the MI for $\alpha \lesssim 1$,
where the area law
for the von Neumann entropy is logarithmically violated
\cite{vodola14,lepori2016}.
{\color{black}These properties are characteristic of phases at $\alpha <1$ that are not are not connected
  with those at $\alpha>1$ \cite{lepori2016}. For instance, they can host gapped edged modes, which
  are absent at $\alpha>1$. Moreover, these phases} are 
not included in the standard classification of the (short-range) topological insulators and superconductors, due to the singularities in the Brillouin zone
from the long-range couplings, see e.g. 
\cite{ludwig2009}.\\
\indent Finally, for completeness we report in Fig. \ref{fp6_bis} the filling
vs the chemical potential for different values of $\alpha$.

\subsection{Long-range hopping and pairing}
\label{sec:LRHopPairing}
Finally, we consider the long-range paired Kitaev model with also long-range hopping:
\begin{eqnarray}
  H & = & - \sum_{i,j=1}^{N_S} \frac{t}{|i-j|_p^{\beta}}
  \left(c^\dagger_i c_{j} + \mathrm{H.c.} \right) -
  \nonumber \\
  &&  - \sum_{i,j=1}^{N_S} \frac{\Delta}{2|i-j|_p^{\alpha}}
    \left(c_j^\dagger c_i^\dagger + c_i c_{j} \right) \, ,
\label{Ham2_ot}
\end{eqnarray}
where we denote by $\beta$ ($\alpha$) the power-law decay exponent
of the hopping (pairing) term.

\begin{figure}[h]
		\includegraphics[width=\columnwidth]{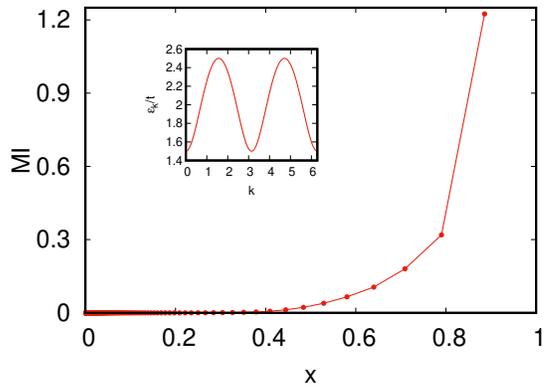}
	        \caption{MI {\it vs} $x$ for the Kitaev chain (\ref{Ham2_ot})
                  with non-local hopping and pairing. The parameters
                  $\alpha = 100$ and $\beta = 0.5$, with
                  $N_S = 1000$ and $\mu = 1.5$, are used. Inset:
                  Bogoliubov quasiparticle spectrum.}
	\label{fp7}
\end{figure}

\begin{figure}[h]
		\includegraphics[width=\columnwidth]{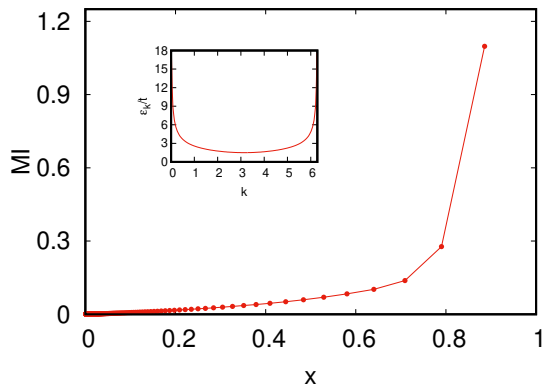}
	        \caption{Same as in Fig. \ref{fp7}, but here $\alpha =\beta =0.5$.}
	\label{fp9}
\end{figure}

We plot in Figs. \ref{fp7} and \ref{fp9} the MI
as a function of $x$ for two different values of the pair $\alpha, \beta$
(with $N_S = 1000$ and $\mu = 1.5$).
In the insets of both figures, the corresponding energy spectrum
is plotted. In Fig. \ref{fp7} we consider
$\alpha =100$ and $\beta = 0.5$, corresponding to a choice
of short-range pairing and
long-range hopping. In this case, it is seen that the
spectrum of the Bogoliubov quasiparticles is regular, not fractal or "zig-zag", like the ones studied
in Sections \ref{sec:fractal} and \ref{subsec:volume}. The MI is vanishing
for small $x$, paralleling the fulfillment of the EE area law, and
conforming to the results obtained for the short-range Kitaev model. Instead,
the case $\alpha = \beta = 0.5$, with both long-range hopping and long-range pairing,
is considered in Fig. \ref{fp9}.
In this case, there are logarithmic violations for the EE area law and, correspondingly, 
we find that the MI is a monotonously
increasing function of the conformal four-point ratio $x$,
similar to the behaviors on massless semi(lines) $\mu \pm 1$, or in the regime $\alpha < 1$, of the model in Eq. \eqref{Ham}.  

\section{Conclusions}
\label{sec:conclusions}
We have studied the mutual information (MI) for several quadratic
fermionic chains with a variety of different long-range hoppings
and superconductive pairings, including the Kitaev model with short- and
long-range pairings and the antipodal model in which the hoppings
connect the most distant sites of the chain.
The MI has been plotted as a function of the
conformal four-point ratio $x$ {\color{black} and of other combinations
  of the physical parameters, such $l/d$ where $l$ and $d$
  are respectively the length of and the distance between the two subsystems.}

The conclusions emerging from the considered examples are the following.
{\it 1)} When the area law is obeyed with at most logarithmic corrections,
the MI is a monotonically increasing function of $x$, going to zero for
$x \to 0$. {\it 2)} {\color{black} $x$ appears as a convenient variable
  for the critical short-range Kitaev model and the non-critical
  long-range Kitaev model, with small values of the exponent $\alpha$
  (where the correlations have a power-law tail),}
  in the sense that different values for the sizes of the system, of the
subsystems and their distance with the same $x$ produces for large
chains the same MI. {\color{black}However $x$ is not in general
  a variable for which MI data collapse, especially for not critical cases, and, as a rule of thumb, when $x$ is no\textcolor{black}t a good variable, other combinations of parameters we tried are not either.}
{\it 3)} When there are non-logarithmic violation of the area law,
the behaviour of MI vs $x$ is no longer monotonic and peaks
appear, with the exception of the antipodal model in which
MI vanishes identically, corresponding to a perfect, maximal volume law for the bipartite EE. {\it 4)} In these non-monotonic cases, $x$ does not
capture the structure of the MI, since parameters with the same $x$ do not have
the same {\color{black} MI} (even though a similar qualitative behaviour is found).

The short-range Kitaev model in the massive regime exhibits
a behaviour such that the
MI is vanishing for small values of $x$, 
\textcolor{black}{noticing however that we are outside of the range of validity for the AdS/CFT prediction (\ref{formula}).} This result is actually independent
from the short or long-range nature of the hopping. When the pairing
becomes strongly long-range ($\alpha<1$), then the MI is no longer vanishing
for small $x$
due to the power-law nature of the correlations induced by the
long-range pairings. We point out that, as expected, for the Kitaev model
we did not find a good agreement between our results for larger values
of $x$ (say $x \gtrsim 1/2$) 
and the AdS/CFT prediction (\ref{formula}), since the latter is valid
in the strong coupling limit. However, in both cases there is
an overall monotonic growth of the MI as a function of the conformal
four-point ratio $x$. For the considered Kitaev models, we showed
the possibility to locate phase transitions using MI.
Keeping the four-point ratio constant and sweeping the phase diagram,
the MI displays distinct peaks at the closings of the gap, as shown
in Figs. \ref{fp5} and \ref{fp6}.

We also analyzed the MI behavior in systems
with volume-like violations of the area law. We observed
the emergence of peaks, and in particular for $x \simeq 0$,
reflecting the structure of shared Bell-pairs between the subsystems.
In fact, violations of the area law imply the formation of
Bell pairs at arbitrary distances, growing with the thermodynamic limit.
From this point of view we can then expect that
in the infinite chain length limit, these peaks get squeezed toward
$x \to 0$ (see Fig. \ref{fig:Antinodal-1}).
To analyze this behaviour we considered models with a controllable
deviation from the perfect (i.e., maximal) volume law exhibited
by the antipodal model. 
For the case of maximal volume law EE
we found a vanishing MI (which can also be interpreted as a
sign of entanglement monogamy)
when the EE decreases the MI increases and peaks appear. The distribution
of the latter is related to the formation of Bell-paired states at
different distances, as it happens in the model with selective hopping
considered in Section \ref{sec:selhop}. A non-monotonic behaviour
of the MI is observed, with features not being
universal in terms of the four-point ratio. We can qualitatively explain this
by the fact that a specific spatial distribution of Bell pairs
cannot be simply captured by conformally invariant quantities.

The presented results show that the MI, even though
not a proper entanglement measure, can be used to extract important
information about the entanglement {\color{black}and quantum correlations} properties and the phases
of a quantum system. The peculiar role of long-range terms,
as intertwined with the possible occurrence of violations of the area law,
have been investigated and shown to produce a variety of interesting
features. It would be certainly very interesting both
to find analytical results for the behaviour of MI as a function
of the conformal four-point ratio for some of the models considered
here and to extend the analysis of the MI to interacting, non-quadratic
models. Indeed, our results show that already for systems that can be mapped to free fermions the MI is not always captured by the existing analytical expressions.
An interesting generalization and application of our study can be the use of MI as a scrambling quantifier (for dynamics under entangling unitary controls),
for both open and closed quantum systems, in condensed matter (spin chains) as well as in high energy physics. This possibility
has been highlighted recently in \cite{deffner2020_1,deffner2020_2}.

\section*{ACKNOWLEDGEMENTS}
The authors are pleased to thank Domenico Giuliano, Erik Tonni,
and Vladimir Korepin
for useful discussions. L. L. acknowledges 
financial support by the PRIN project number 20177SL7HC,
financed by the Italian Ministry of education and research.  L. L. also acknowledges financial support by the Qombs Project [FET Flagship on Quantum Technologies grant n. 820419].
F. F. acknowledges support from the Croatian Science Foundation (HrZZ) Projects No. IP--2016--6--3347 and IP--2019--4--3321, as well as from from the QuantiXLie Center of Excellence, a project co--financed by the Croatian Government and European Union through the European Regional Development Fund -- the Competitiveness and Cohesion (Grant No. KK.01.1.1.01.0004).

\end{document}